\documentclass[aps,prb,twocolumn]{revtex4}
\usepackage{graphicx}
\usepackage{natbib}

\begin{document}

\title{Ion and polymer dynamics in polymer electrolytes PPO-LiClO$_4$:
\\II.\ $^2$H and $^7$Li NMR stimulated-echo experiments}

\author{M.\ Vogel}
\author{T.\ Torbr\"ugge}
\affiliation{Institut f\"ur Physikalische Chemie, Westf\"alische
Wilhelms-Universit\"at M\"unster, Corrensstr.\ 30/36, 48149
M\"unster, Germany}

\date{\today}

\begin{abstract}
We use $^2$H NMR stimulated-echo spectroscopy to measure two-time correlation functions
characterizing the polymer segmental motion in polymer electrolytes PPO-LiClO$_4$ near
the glass transition temperature $T_g$. To investigate effects of the salt on the polymer
dynamics, we compare results for different ether oxygen to lithium ratios, namely, 6:1,
15:1, 30:1 and $\infty$. For all compositions, we find nonexponential correlation
functions, which can be described by a Kohlrausch function. The mean correlation times
show quantitatively that an increase of the salt concentration results in a strong
slowing down of the segmental motion. Consistently, for the high 6:1 salt concentration,
a high apparent activation energy $E_a\!=\!4.1\mathrm{\,eV}$ characterizes the
temperature dependence of the mean correlation times at $T_g\!<\!T\!\lesssim\!1.1\,T_g$,
while smaller values $E_a\!\approx\!2.5\mathrm{\,eV}$ are observed for moderate salt
contents. The correlation functions are most nonexponential for 15:1 PPO-LiClO$_4$,
whereas the stretching is reduced for higher and lower salt concentrations. This finding
implies that the local environments of the polymer segments are most diverse for
intermediate salt contents, and, hence, the spatial distribution of the salt is most
heterogeneous. To study the mechanisms of the segmental reorientation, we exploit that
the angular resolution of $^2$H NMR stimulated-echo experiments depends on the length of
the evolution time $t_p$. A similar dependence of the correlation functions on the value
of $t_p$ in the presence and in the absence of ions indicates that addition of salt
hardly affects the reorientational mechanism. For all compositions, mean jump angles of
about 15$^\circ$ characterize the segmental reorientation. In addition, comparison of
results from $^2$H and $^7$Li NMR stimulated-echo experiments suggests a coupling of ion
and polymer dynamics in 15:1 PPO-LiClO$_4$.
\end{abstract}
\maketitle

\section{Introduction}

The glass transition of polymer melts is of tremendous interest from the viewpoints of
fundamental and applied science. This importance triggered a huge amount of studies,
which established that the glass transition phenomenon of polymers is related to the
$\alpha$ process.\cite{McCrum,Ferry,Spiess} An improvement of our understanding of the
glass transition thus requires detailed characterization of the microscopic dynamics
underlying the $\alpha$ process. In particular, it is important to determine both the
time scale and the geometry of the molecular dynamics.

Various methods were used to measure the correlation time of the $\alpha$ process,
$\tau_{\alpha}$. For polymers, a striking feature of the $\alpha$ process is a
non-Arrhenius temperature dependence. In most cases, the Vogel-Fulcher-Tammann equation,
\begin{math}
\tau_{\alpha}(T)\!=\!\tau_0\exp\left[B/(T\!-\!T_0)\right]
\end{math},
enables a satisfactory interpolation of the data, suggesting a divergence of the
correlation time at a temperature $T_0$ not too far below the glass transition
temperature $T_g$, i.e., $T_0\!\approx\!T_g\!-\!50\mathrm{\,K}$.\cite{Spiess} Hence, high
apparent activation energies $E_a$ characterize the temperature dependence of the
$\alpha$ process close to $T_g$.\cite{Spiess} Another intriguing property of the $\alpha$
process is a nonexponential loss of correlation, which is often well described by the
Kohlrausch-Williams-Watts (KWW) function,
\begin{math}
\exp[-(t/\tau)^\beta]
\end{math}.
By contrast, few experimental techniques yield detailed insights into the geometry of the
molecular motion associated with the $\alpha$ process.

2D NMR provides straightforward access to both the time scale and the geometry of
molecular reorientation in polymeric and molecular glass-forming
liquids.\cite{Spiess,Roessler,Boehmer,Duer,Reichert} 2D NMR experiments can be performed
in the frequency domain\cite{Wefing,Kaufmann,Pschorn,Schaefer} and in the time
domain.\cite{SpiessTD,Diehl,RoEi,Geil,Hinze,BoHi,Tracht} The latter approach is also
called stimulated-echo (SE) spectroscopy. Concerning the geometry of polymer dynamics, 2D
NMR studies ruled out the existence of well-defined jump angles $\gamma$ for the
rotational jumps associated with the $\alpha$
process.\cite{Spiess,Boehmer,Reichert,Wefing,Kaufmann,Pschorn,Schaefer,Tracht} Instead,
results from 2D and 3D NMR showed that the reorientation is comprised of rotational jumps
about angles $\gamma\!=\!10\!-\!40^{\circ}$ between which the polymer segments show
isotropic rotational diffusion.\cite{Schaefer,Tracht,Leisen,Heuer,Kuebler}

We carry out $^2$H NMR SE experiments to investigate the segmental motion of
poly(propylene oxide) (PPO) containing various amounts of the salt LiClO$_4$. Thus, we
focus on polymer electrolytes, which are promising candidates for applications in energy
technologies.\cite{Gray} Since the use of polymer electrolytes is still limited by the
achievable electric conductivities, there is considerable interest to speed up the ionic
diffusion. It has been established that segmental motion of the host polymer facilitates
the ionic migration,\cite{Armand,Ratner} yet the understanding of this coupling is
incomplete. In dielectric spectroscopy, the ionic diffusion coefficient and the
relaxation rate of polymer segments in contact with salt were found to show a linear
relationship,\cite{Furukawa} suggesting that segmental motion triggers the elementary
steps of the charge transport.

Polymer electrolytes PPO-LiClO$_4$ have attracted much attention because of high ionic
conductivities together with their amorphous nature.\cite{Armand} When studied as a
function of the salt concentration, the electric conductivity $\sigma_{dc}$ at room
temperature shows a broad maximum at ether oxygen to lithium ratios, O:Li, ranging from
about 10:1 to 30:1.\cite{Furukawa,McLin} For such intermediate salt concentrations,
differential scanning calorimetry showed the existence of two glass transition
steps,\cite{Moacanin,Vachon_1,Vachon_2} implying a liquid-liquid phase separation into
salt-rich and salt-depleted regions. An inhomogeneous structure was also assumed to
explain results from dielectric and photon correlation spectroscopies for intermediate
salt concentrations,\cite{Furukawa,Bergman} which give evidence for the presence of two
processes related to fast and slow polymer dynamics. By contrast, the existence of
salt-rich and salt-depleted microphases was challenged in neutron scattering work on
mixtures PPO-LiClO$_4$.\cite{Carlsson_1,Carlsson_2} Finally, it was proposed that a
structure model of 16:1 PPO-LiClO$_4$ resulting from reverse Monte-Carlo simulations of
diffraction data enables reconciliation of the apparent discrepancies.\cite{Carlsson_3}
It does feature salt-rich and salt-depleted regions, but the structural heterogeneities
occur on a small length scale of $\mathrm{1\,nm}$.

In NMR work on polymer electrolytes, the ion dynamics was studied in some
detail,\cite{Roux,Chung_1,Chung_2,Adamic,Fan,Donoso,Forsyth} e.g., ionic self diffusion
coefficients were determined using field-gradient techniques,\cite{Vincent,Arumugam,Ward}
while investigations of the polymer dynamics were restricted to analysis of the
spin-lattice relaxation.\cite{Ward,Roux,Donoso} Only recently, modern NMR methods were
employed to study the segmental motion in organic-inorganic hybrid
electrolytes.\cite{Kao} We use various $^2$H and $^7$Li NMR techniques for a detailed
characterization of the polymer and the ion dynamics in polymer electrolytes
PPO-LiClO$_4$. Our goals are twofold. From the viewpoint of fundamental science, we
ascertain how the presence of salt affects the polymer dynamics. In this way, we tackle
the question how structural heterogeneities introduced by the addition of salt affect the
glass transition of polymers. From the viewpoint of applied science, we investigate how
the polymer assists the ionic diffusion so as to give support to the development of new
materials. For example, we compare $^2$H and $^7$Li NMR data to study the coupling of ion
and polymer dynamics.

Very recently, we presented results from $^2$H and $^7$Li NMR line-shape analysis for
polymer electrolytes PPO-LiClO$_4$.\cite{Vogel} The $^2$H NMR spectra indicate that
addition of salt slows down the segmental motion. Moreover, the line shape shows that
broad distributions of correlation times $G(\log \tau)$ govern the polymer dynamics.
These dynamical heterogeneities are particularly prominent for intermediate salt
concentrations. While such line-shape analysis provides valuable insights on a
qualitative level, the possibilities for a quantification are limited.

Here, we exploit that $^2$H NMR SE spectroscopy enables measurement of two-time
correlation functions that provide quantitative information about polymer segmental
motion on a microscopic level.\cite{Spiess} In this way, we continue to investigate to
which extent presence of salt affects the $\alpha$ process of polymer melts. While
previous $^2$H NMR SE studies focussed on molecules featuring a single deuteron species,
two deuteron species, which exhibit diverse spin-lattice relaxation (SLR) and spin-spin
relaxtion (SSR), contribute to the correlation functions in our case. Therefore, it is
necessary to ascertain the information content of this method in such situation. We will
demonstrate that, in the presence of two deuteron species, it is necessary to use
appropriate experimental parameters, when intending to correct the measured data for
relaxation effects, which is highly desirable in order to extend the time window of the
experiment.

\section{Theory}

\subsection{Basics of $^2$H and $^7$Li NMR}\label{NMR}

Solid-state $^2$H NMR probes the quadrupolar precession frequency $\omega_Q$, which is
determined by the orientation of the quadrupolar coupling tensor with respect to the
external static magnetic flux density $\mathbf{B_0}$. This tensor describes the
interaction of the nuclear quadrupole moment with the electric field gradient at the
nuclear site. The monomeric unit of the studied PPO, [CD$_2$-CD(CD$_3$)-O]$_n$, features
three deuterons in the backbone and three deuterons in the methyl group, which will be
denoted as B deuterons and M deuterons, respectively. For both deuteron species, the
coupling tensor is approximately axially symmetric.\cite{Vogel} Then, $\omega_Q$ is given
by\cite{Spiess}
\begin{equation}\label{omega}
\omega_Q(\theta)=\pm \frac{\delta_{y}}{2}\,(3\cos^2\theta_{y} -1)
\end{equation}
Here, $y\!=\!B,M$ refers to the different deuteron species. For the B deuterons,
$\theta_{B}$ specifies the angle between the axis of the C-D bond and $\mathbf{B_0}$ and
the anisotropy parameter amounts to
$\delta_B\!\approx\!2\pi\!\times\!\mathrm{120\,kHz}$.\cite{Vogel} For the M deuterons,
the coupling tensor is preaveraged due to fast threefold jumps of the methyl group at the
studied temperatures. Consequently, $\theta_{M}$ is the angle between the threefold axis
and $\mathbf{B_0}$ and the anisotropy parameter is reduced by a factor of three,
$\delta_{M} \!=\!\delta_{B}/3$.\cite{Spiess} The two signs in Eq.\ (\ref{omega})
correspond to the two allowed transitions between the three Zeeman levels of the
$I\!=\!1$ nucleus. We see that, for both deuteron species, the quadrupolar precession
frequency is intimately linked to the orientation of well defined structural units and,
hence, segmental motion renders $\omega_Q$ time dependent, which is the basis of $^2$H
NMR SE spectroscopy.\cite{Spiess,Roessler}

In solid-state $^7$Li NMR ($I\!=\!3/2$), the quadrupolar interaction affects the
frequencies of the satellite transitions $\pm3/2\!\leftrightarrow\!\pm1/2$, while that of
the central transition $1/2\!\leftrightarrow\!-1/2$ is unchanged. For polymer
electrolytes, diverse lithium ionic environments lead to a variety of shapes of the
quadrupolar coupling tensor so that $^7$Li NMR spectra exhibit an unstructured satellite
component.\cite{Chung_1,Vogel} Analysis of $^7$Li NMR SE experiments will be
straightforward if the time dependence $\omega_Q(t)$ is solely due to lithium ionic
motion. This condition is approximately met in applications on solid-state electrolytes,
i.e., when the lithium ionic diffusion occurs in a static
matrix.\cite{BoehmerLi_1,BoehmerLi_2} For polymer electrolytes, the lithium environments
change due to ion \emph{and} polymer dynamics, occurring on comparable time scales.
Consequently, both these dynamic processes can render the electric field gradient at the
nuclear site and, hence, $\omega_Q$ time dependent.\cite{Zax} Thus, when $^7$Li NMR is
used to investigate charge transport in polymer electrolytes, it is a priori not clear to
which extent the lithium ionic diffusion and the rearrangement of the neighboring polymer
chains are probed.

\subsection{$^2$H NMR stimulated-echo spectroscopy}\label{THEO}

$^2$H NMR SE spectroscopy proved a powerful tool to study molecular dynamics with
correlation times in the range of $10\,\mathrm{\mu
s}\!\leq\!\tau\!\leq\!1\mathrm{\,s}$.\cite{Boehmer,SpiessTD,Diehl,RoEi,Geil,Hinze,BoHi,Fujara_1,Fujara_2,Vogel_1,Vogel_2,Jeffrey}
In this experiment, the frequency $\omega_Q$ is probed twice during two short evolution
times $t_p\!\ll\tau$, which embed a long mixing time $t_m\!\approx\tau$. In detail,
appropriate three-pulse sequences are used to generate a stimulated
echo.\cite{Spiess,Duer,Roessler,Boehmer,Reichert} Evaluating the height of this echo for
various mixing times $t_m$ and constant evolution time $t_p$, it is possible to measure
two-time correlation functions. Neglecting relaxation effects, one obtains depending on
the pulse lengths and phases in the three-pulse
sequence\cite{Spiess,Duer,Roessler,Boehmer,Reichert}
\begin{equation}\label{SS}
E_2^{SS}(t_m;t_p)=\langle\, \sin[\,\omega_Q(0)t_p]\cdot\sin[\,\omega_Q(t_m)t_p]\,\rangle,
\end{equation}
\begin{equation}\label{CC}
E_2^{CC}(t_m;t_p)=\langle\, \cos[\,\omega_Q(0)t_p]\cdot\cos[\,\omega_Q(t_m)t_p]\,\rangle.
\end{equation}
Here, the brackets $\langle\dots\rangle$ denote the ensemble average, including the
powder average for our isotropic distribution of molecular orientations.\cite{34} We see
that the frequencies $\omega_Q(0)$ and $\omega_Q(t_m)$ are correlated via sine or cosine
functions. In general, molecular reorientation during the mixing time leads to
$\omega_Q(0)\!\neq\!\omega_Q(t_m)$ and, hence, to decays of $E_2^{SS}(t_m)$ and
$E_2^{CC}(t_m)$. Normalizing these correlation functions, we obtain ($x\!=\!SS,CC$)
\begin{equation}\label{F}
F_2^{x}(t_m;t_p)=\frac{E_2^{x}(t_m;t_p)}{E_2^{x}(0;t_p)}
\end{equation}
to which we will refer as sin-sin and cos-cos correlation functions in the following.

The dependence of $F_2^{x}(t_m;t_p)$ on the evolution time provides access to the jump
angles of molecular rotational
jumps.\cite{Hinze,BoHi,Geil,Tracht,Jeffrey,Vogel_1,Vogel_2} This approach exploits that
the angular resolution of the experiment depends on the length of the evolution time
$t_p$. When the quadrupolar coupling tensor is axially symmetric, the sin-sin correlation
function for $t_p\!\rightarrow\!0$ yields the rotational correlation function of the
second Legendre polynomial, $F_2(t_m)$:
\begin{eqnarray}\label{F2}
F_2^{SS}(t_m;t_p\!\rightarrow\! 0)\propto
t_p^2\,\langle\,\omega_Q(0)\,\omega_Q(t_m)\,\rangle \propto \nonumber \\ \propto \langle
P_2[\cos(\theta(0)]P_2[\cos(\theta(t_m)]\rangle \propto F_2(t_m).
\end{eqnarray}
On extension of the evolution time, smaller and smaller changes of the resonance
frequency, $|\omega_Q(t_m)\!-\!\omega_Q(0)|$, are sufficient to observe a complete decay
of $F_2^{x}(t_m;t_p)$, i.e., isotropic redistribution of the molecular orientations is no
longer required. In the limit $t_p\!\rightarrow\!\infty$,
\begin{equation}\label{JCF}
F_2^{x}(t_m;t_p\!\rightarrow\!\infty)= \langle \delta[\,\omega_Q(t_m)\!-\!\omega_Q(0)]
\rangle \equiv F_2^J(t_m)
\end{equation}
and, hence, the jump correlation function $F_2^J(t_m)$ is measured, which characterizes
the elementary steps of the reorientation process. The mean jump angle $\gamma$ of the
molecular rotational jumps can be determined by comparison of the correlation times
$\tau_C$ and $\tau_J$ characterizing the decays of $F_2(t_m)$ and $F_2^J(t_m)$,
respectively,\cite{Anderson}
\begin{equation}\label{gamma}
\frac{\tau_J}{\tau_C}=\frac{3}{2}\sin^2\gamma.
\end{equation}
Analysis of the complete dependence of the correlation time on the evolution time,
$\tau(t_p)$, provides further information about the distribution of jump
angles.\cite{Hinze,BoHi,Tracht}

Here, the existence of two deuteron species complicates the analysis of sin-sin and
cos-cos correlation functions. Thus, prior to a presentation of experimental results, it
is important to determine the interpretation of these correlation functions in such
situation. In the first step, we will continue to neglect relaxation effects and study
consequences of the different anisotropy parameters $\delta_B\!=\!3\delta_M$. In the
second step, we will investigate how the interpretation of the correlation functions is
affected by the fact that the deuteron species exhibit different spin-lattice and
spin-spin relaxation times, $T_1$ and $T_2$, respectively.

In the first step, the non-normalized correlation functions can be written as the
following sum of contributions from the B and the M deuterons
\begin{eqnarray}\label{E}
E_2^x(t_m;t_p)&=&\sum_y m_y e_{2,y}^x(t_m;t_p) \nonumber\\
e_{2,y}^x(t_m;t_p)&\!=\!&
\left[a_y^x(t_p)\!-\!z_y^x(t_p)\right]\phi_y^x(t_m;t_p)\!+\!z_y^x(t_p) \:\:\:\:\:
\end{eqnarray}
Here, we consider that $e_{2,y}^x(t_m;t_p)$ decays from an initial correlation
$a_y^x(t_p)$ to a residual correlation $z_y^x(t_p)$ and, hence, the correlation functions
$\phi_y^x(t_m;t_p)$ are normalized according to $\phi_y^x(0)\!=\!1$ and
$\phi_y^x(\infty)\!=\!0$. The molar ratios of the deuteron species, $m_y\!=\!1/2$, are
proportional to their contributions to the equilibrium magnetization. The initial state
values of the correlation functions are determined by ($\xi\!=\!\sin,\cos$)
\begin{equation}\label{A}
e_{2,y}^x(0;t_p)=a_y^x(t_p)=\langle \xi^2[\,\omega_Q t_p]\rangle_y
\end{equation}
where $\langle\dots\rangle_y$ denotes the powder average for the respective species. The
final state values depend on the geometry of the reorientation
process.\cite{Fujara_1,Fujara_2} For an isotropic motion, the frequencies $\omega_Q(0)$
and $\omega_Q(t_m)$ are uncorrelated in the limit $t_m\!\gg\!\tau$, leading to
\begin{equation}\label{Z}
e_{2,y}^x(\infty;t_p)=z_y^x(t_p)=\langle \xi[\,\omega_Q t_p]\rangle_y^2
\end{equation}

\begin{figure}
\includegraphics[angle=0,width=6.5cm]{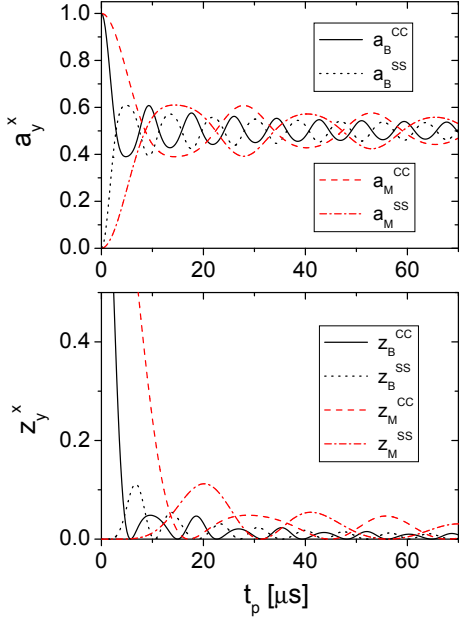}
\caption{Initial correlations $a_y^x(t_p)$ and residual correlations $z_y^x(t_p)$ for an
isotropic motion, see Eq.\ (\ref{E}). The values were calculated for anisotropy
parameters $\delta_B\!=\!2\pi\!\times\!120\mathrm{\,kHz}$ and
$\delta_M\!=\!2\pi\!\times\!40\mathrm{\,kHz}$.}\label{AZ}
\end{figure}

In Fig.\ \ref{AZ}, we show theoretical values of the initial and residual correlations
$a_y^x(t_p)$ and $z_y^x(t_p)$, respectively. For the calculation, we used typical
anisotropy parameters $\delta_B\!=\!2\pi\!\times\!120\mathrm{\,kHz}$ and
$\delta_M\!=\!2\pi\!\times\!40\mathrm{\,kHz}$.\cite{Vogel} Due to the periodicity of the
sine and cosine functions, all the quantities show an oscillatory behavior with a
frequency proportional to the anisotropy parameter. Thus, as a consequence of
$\delta_B\!\neq\!\delta_M$, the ratios $a_B^x/a_M^x$ are functions of the evolution time,
indicating that the relative contributions of the two deuteron species to the correlation
functions depend on the value of $t_p$.

SLR and SSR limit the lengths of the interpulse delays in SE experiments. To consider
relaxation effects in the second step, we modify Eq.\ (\ref{E}) according to
\begin{equation}\label{R}
{\cal E}_2^x(t_m;t_p)\!=\!\sum_y m_y e_{2,y}^x(t_m;t_p)R_{1,y}(t_m)R_{2,y}(2t_p)
\end{equation}
While $R_{1,y}(t_m)$ describes the decrease of the echo height due to SLR during the
mixing time, $R_{2,y}(2t_p)$ quantifies the loss of signal due to SSR during the
evolution times, i.e., $R_{i,y}(0)\!=\!1$ and $R_{i,y}(\infty)\!=\!0$ ($i\!=\!1,2$). In
analogy with Eq.~(\ref{F}), we define ${\cal F}_2^x(t_m;t_p)\!=\!{\cal
E}_2^x(t_m;t_p)/{\cal E}_2^x(0;t_p)$. The symbols ${\cal E}_2^x$ and ${\cal F}_2^x$ are
used to stress that the correlation functions are affected by relaxation effects. As a
consequence of fast methyl group rotation at the studied temperatures, the SLR times of
the M deuterons are shorter than that of the B deuterons,
$T_{1,M}\!\ll\!T_{1,B}$.\cite{Vogel} Furthermore, the deuteron species exhibit different
SSR and, hence, the ratio of the respective contributions to the correlation functions
increasingly deviates from the theoretical value $a_B^x(t_p)/a_M^x(t_p)$, when the
evolution time is increased, see Appendix.

The time window of the experiment is limited in particular by the fast SLR of the M
deuterons at the studied temperatures, $T_{1,M}\!\approx\!10\mathrm{\,ms}$. Thus,
correction for SLR is desirable so as to extend the time window. In $^2$H NMR, such
correction is not possible for the sin-sin correlation functions, since the decay of
Alignment order, which is present during the mixing time of this experiment, is not
accessible in an independent measurement so that the damping due to SLR is unknown. By
contrast, for the cos-cos correlation functions, the decay of Zeeman order, which exists
during the mixing time of this measurement, can be determined in a regular SLR
experiment. In previous $^2$H NMR SE studies, it thus proved useful to focus on the
cos-cos correlation function and to divide ${\cal E}_2^{CC}$ by the SLR function
$R_1(t_m)$ to remove SLR effects.

Following an analogous approach, we analyze the correlation functions
\begin{equation}\label{FTP}
F_2^{*}(t_m;t_p)=\frac{{\cal E}_2^{CC}(t_m;t_p)}{\sum_y
m_y\,a_y^{CC}(t_p)\,R_{1,y}(t_m)R_{2,y}(2t_p)}
\end{equation}
Here, the denominator can be regarded as an effective SLR function. This function takes
into account that the contributions of the B and the M deuterons to the cos-cos
correlation function are weighted with the initial correlations $a_y^{CC}(t_p)$.
Moreover, it includes that the relative contributions of the two species change due to
diverse SSR, as is evident from the factors $R_{2,y}(2t_p)$.

In the Appendix, we discuss whether this approach enables correction for SLR. Our
considerations show that, when two (or more) deuteron species contribute to the cos-cos
correlation functions, elimination of relaxation effects is successful only when three
conditions are met: (i) Both species exhibit the same dynamical behavior, i.e.,
$\phi_B^{CC}(t_m;t_p)\!=\!\phi_M^{CC}(t_m;t_p)$, which is fulfilled to a good
approximation in our case, see Sec.\ \ref{TP}. (ii) Appropriate evolution times are used
so that $z_B^{CC}\!=\!z_M^{CC}\!\approx\!0$. (iii) The data are corrected utilizing the
effective SLR function defined in Eq.\ (\ref{FTP}). The experimental determination of
this function is described in the Appendix. In the following, we will mostly focus on
evolution times, for which correction for SLR is possible, i.e.,
$F_2^{*}(t_m;t_p)\!\approx\!F_2^{CC}(t_m;t_p)$. In these cases, the corrected data will
be referred to as $F_2^{CC}$, while we will use $F_2^{*}$ to denote situations, when
elimination of SLR effects is not successful.

\subsection{$^7$Li NMR stimulated-echo spectroscopy}

Since the methodologies of $^2$H and $^7$Li NMR SE spectroscopies are very similar, we
restrict ourselves to point out some differences. In $^7$Li NMR, the use of appropriate
three-pulse sequences allows one to measure $F_2^{SS}$, whereas $F_2^{CC}$ is not
accessible for $I\!=\!3/2$.\cite{BoJMR}  Moreover, the shape of the quadrupolar coupling
tensor is well defined in $^2$H NMR, while it is subject to a broad distribution in
$^7$Li NMR, leading to a damping of the oscillatory behavior of the initial and final
correlations, in particular, $z^{SS}\!=\!0$ for sufficiently long evolution times.
Furthermore, in our case, $^7$Li SLR is slow so that correction of the $^7$Li NMR
correlation functions for relaxation effects is not necessary. Finally, we again
emphasize that both ion and polymer dynamics can contribute to the decays of $^7$Li NMR
correlation functions and, hence, their interpretation is not as straightforward as for
the $^2$H NMR analogs.

\section{Experiment}\label{exp}

In addition to neat PPO, we study the polymer electrolytes 30:1 PPO-LiClO$_4$, 15:1
PPO-LiClO$_4$ and 6:1 PPO-LiClO$_4$. In the following, we will refer to these samples as
PPO, 30:1 PPO, 15:1 PPO and 6:1 PPO. In all samples, we use a deuterated polymer
[CD$_2$-CD(CD$_3$)-O]$_\mathrm{n}$ with a molecular weight
$M_w\!=\!\mathrm{6510\,g/mol}$. Strictly speaking, the polymer is poly(propylene glycol),
having terminal OH groups. The sample preparation was described in our previous
work.\cite{Vogel} In DSC experiments,\cite{Vogel} we found $T_g\!=\!\mathrm{203\pm2\,K}$
for both PPO and 30:1 PPO and $T_g\!=\!\mathrm{272\pm2\,K}$ for 6:1 PPO. For the
intermediate 15:1 composition, we observed a broad glass transition step ranging from
about $\mathrm{205\,K}$ to $\mathrm{245\,K}$. In the literature, two glass transition
temperatures $T_{g,1}\!\approx\!\mathrm{208\,K}$ and $T_{g,2}\!\approx\!\mathrm{240\,K}$
were reported for 16:1 PPO-LiClO$_4$.\cite{Vachon_1}

The $^2$H NMR experiments were performed on Bruker DSX 400 and DSX 500 spectrometers
working at Larmor frequencies $\omega_0/2\pi$ of $\mathrm{61.4\,MHz}$ and
$\mathrm{76.8\,MHz}$, respectively. Two Bruker probes were used to apply the radio
frequency pulses, resulting in $90^\circ$ pulse lengths of $\mathrm{2.2\,\mu s}$ and
$\mathrm{3.5\,\mu s}$. Comparing results for different setups, we determined that the
results depend neither on the magnetic field strength nor on the probe, see Sec.\
\ref{TD}. The $^7$Li NMR measurements were carried out on the DSX 500 spectrometer
working at a Larmor frequency of $\mathrm{194.4\,MHz}$. In these experiments, the
$90^\circ$ pulse length amounted to $\mathrm{2.5\,\mu s}$. Further experimental details
can be found in our previous work.\cite{Vogel}

Appropriate three-pulse sequences $P_1$ - $t_p$ - $P_2$ - $t_m$ - $P_3$ - $t$ are
sufficient to measure sin-sin and cos-cos correlation functions for sufficiently large
evolution times. However, for short evolution times, the stimulated echo, which occurs at
$t\!=\!t_p$, vanishes in the dead time of the receiver, following the excitation
pulses.\cite{Spiess,Roessler} This problem can be overcome, when a fourth pulse is added
to refocus the stimulated echo outside the dead time, leading to four-pulse sequences
$P_1$ - $t_p$ - $P_2$ - $t_m$ - $P_3$ - $\Delta$ - $P_4$.\cite{Spiess,Roessler} We use
three-pulse and four-pulse sequences to measure correlation functions for large
($t_p\!=\!60\mathrm{\,\mu s}$) and small ($t_p\!=\!2\mathrm{\,\mu s}$) evolution times,
respectively. When we study the dependence of the cos-cos correlation functions on the
evolution time, we apply a four-pulse sequence for all values of $t_p$. In all
experiments, we use suitable phase cycles to eliminate unwanted signal
contributions.\cite{Schaefer_PC}

\section{Results}

\subsection{Temperature dependent $^2$H NMR correlation functions}\label{TD}

\begin{figure}
\includegraphics[angle=270,width=8.2cm]{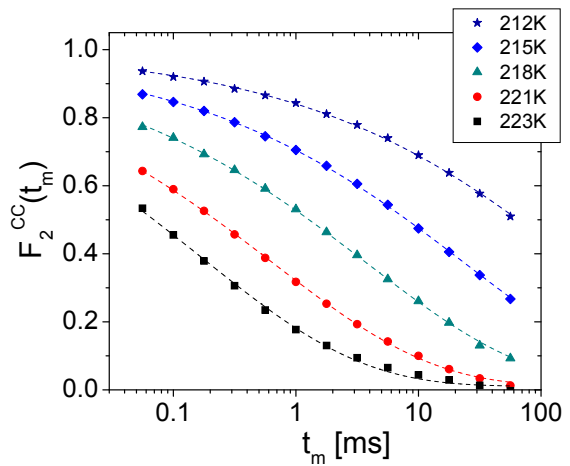}
\caption{Correlation functions $F_2^{CC}(t_m;t_p\!=\!60\mathrm{\,\mu s})$ for neat PPO at
various temperatures. The dashed lines are fits to Eq.\ (\ref{KWW}) with $\beta\!=\!0.33$
and $C\!=\!0.01$.}\label{PPO}
\end{figure}

\begin{figure}
\includegraphics[angle=270,width=8.2cm]{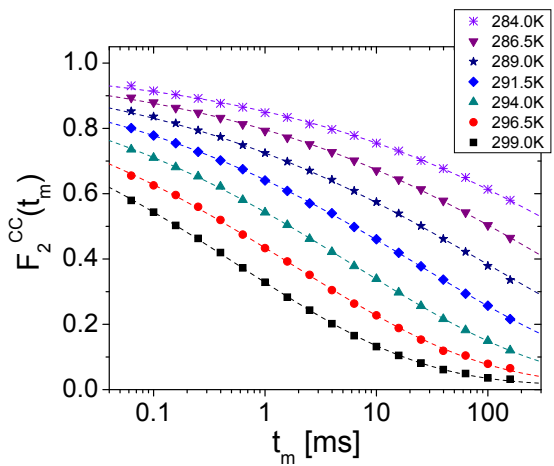}
\caption{Correlation functions $F_2^{CC}(t_m;t_p\!=\!60\mathrm{\,\mu s})$ for 6:1
PPO-LiClO$_4$ at various temperatures. The dashed lines are fits to Eq.\ (\ref{KWW}) with
$\beta\!=\!0.24$ and $C\!=\!0.01$.}\label{PPO6}
\end{figure}

In Figs.\ \ref{PPO} and \ref{PPO6}, we show temperature dependent correlation functions
$F_2^{CC}(t_m;t_p\!=\!60\mathrm{\,\mu s})$ for PPO and 6:1 PPO. Due to the finite pulse
lengths this value of $t_p$ corresponds to an effective evolution time of about
$62\mathrm{\,\mu s}$, see Sec.~\ref{TP}, so that the residual correlation for both
deuteron species approximately vanishes and, hence, correction for SLR is possible.
Furthermore, we will show in Sec.\ \ref{TP} that, for this large evolution time,
$F_2^{CC}(t_m;t_p)$ approximates the jump correlation function $F_2^J(t_m)$, see
Eq.~(\ref{JCF}). In Figs.\ \ref{PPO} and \ref{PPO6}, it is evident that both materials
exhibit nonexponential correlation functions. Moreover, for the 6:1 polymer electrolyte,
the loss of correlation occurs at much higher temperatures in the time window of the
experiment, indicating that the presence of salt slows down the polymer segmental motion.

\begin{figure}
\includegraphics[angle=270,width=8.2cm]{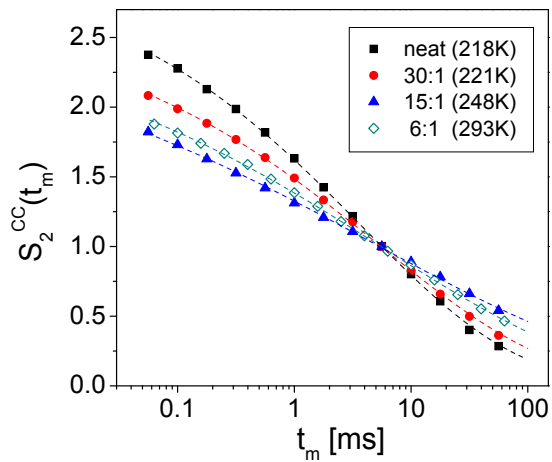}
\caption{Scaled cos-cos correlation functions for PPO, 30:1 PPO-LiClO$_4$, 15:1
PPO-LiClO$_4$ and 6:1 PPO-LiClO$_4$ ($t_p\!=\!60\mathrm{\,\mu s}$). The temperatures are
indicated. They are chosen so as to compare correlation functions with similar time
constants $\tau\!=\!3.8\!-5.8\mathrm{\,ms}$. For the sake of comparison, the data was
scaled according to $S_2^{CC}(t_m)\!=\!F_2^{CC}(t_m)/F_2^{CC}(t_0)$ where
$t_0\!=\!5.6\mathrm{\,ms}\!\approx\!\tau$. The dashed lines are interpolations with a
function proportional to Eq.\ (\ref{KWW}). The used stretching parameters are
$\beta\!=\!0.33$ for PPO, $\beta\!=\!0.3$ for 30:1 PPO-LiClO$_4$, $\beta\!=\!0.2$ for
15:1 PPO-LiClO$_4$ and $\beta\!=\!0.24$ for 6:1 PPO-LiClO$_4$.}\label{VER}
\end{figure}

To study how the presence of salt affects the nonexponentiality of the polymer dynamics,
we compare correlation functions $F_2^{CC}(t_m;t_p\!=\!60\mathrm{\,\mu s})$ for all
studied materials in Fig.\ \ref{VER}. The respective temperatures were chosen so as to
consider correlation functions with comparable time constants. For the sake of
comparison, we actually show scaled correlation functions $S_2^{CC}(t_m)$. We see that
the nonexponentiality of the polymer dynamics is not a linear function of the salt
content, but it is most pronounced for 15:1 PPO. Assuming that the rate of the segmental
motion reflects the local salt concentration, these findings suggest that structural
heterogeneities relating to the spatial salt distribution are most pronounced for the
intermediate 15:1 composition, while the structure is more homogeneous for higher and
lower salt contents.

\begin{figure}
\includegraphics[angle=0,width=8.7cm]{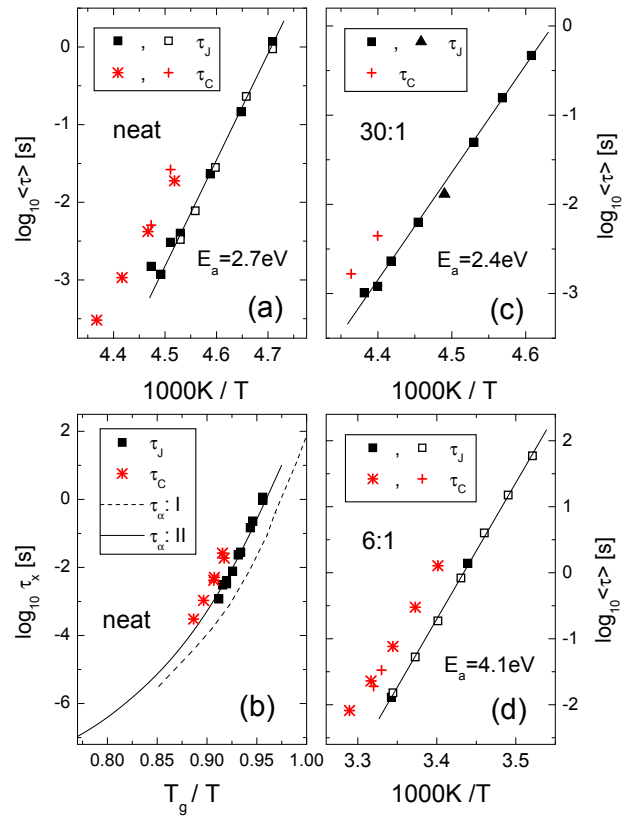}
\caption{Mean correlation times for (a) PPO, (c) 30:1 PPO-LiClO$_4$ and (d) 6:1
PPO-LiClO$_4$. The time constants $\langle\tau_C\!\rangle$ and $\langle\tau_J\!\rangle$
result from interpolations of $F_2^{SS}(t_m;t_p\!=\!2\mathrm{\,\mu s})$ and
$F_2^{CC}(t_m;t_p\!=\!60\mathrm{\,\mu s})$ with Eq.~(\ref{KWW}), respectively. While the
time constants shown as open squares and crosses were obtained for
$\omega_0/(2\pi)\!=\!61.4\mathrm{\,MHz}$, that displayed as solid squares and stars were
determined for $\omega_0/(2\pi)\!=\!76.8\mathrm{\,MHz}$. For 30:1 PPO-LiClO$_4$, the
solid triangle is a data point resulting from experiments using a four-pulse rather than
a three-pulse sequence. In panel (b), we compare our results for PPO with correlation
times $\tau_{\alpha}$ from dielectric spectroscopy (I: Ref.\ \onlinecite{Leon}, II: Ref.\
\onlinecite{Furukawa}). The correlation times are shown on a reduced $T_g/T$ scale to
remove effects due to different glass transition temperatures, $T_g\!=\!203\mathrm{\,K}$
and $T_g\!=\!198\mathrm{\,K}$.\cite{Furukawa,Leon} The latter difference reflects the
fact that polymers with a smaller molecular weight $M_w\!=\!\mathrm{4000\,g/mol}$ were
investigated in the previous approaches.\cite{Furukawa,Leon}}\label{TAU}
\end{figure}

For a quantitative analysis of the correlation functions
$F_2^{CC}(t_m;t_p\!=\!60\mathrm{\,\mu s})$, we fit the data for all compositions and
temperatures with a modified KWW function
\begin{equation}\label{KWW}
(1-C)\exp\left[-\left(\frac{t}{\tau}\right)^\beta\right]+C
\end{equation}
At high temperatures, when the decay to the final state value can be observed in our time
window, we obtain small residual correlations $C\!=\!0.01\!\pm\!0.01$, in harmony with
the theoretical value for this evolution time, see Fig.~\ref{AZ}. Therefore, we keep this
value fixed in the fits for the lower temperatures. For neither of the materials, we find
evidence for a systematic temperature dependence of the stretching parameter $\beta$.
Specifically, we observe values $\beta\!=\!0.33\!\pm\!0.02$ for PPO,
$\beta\!=\!0.30\!\pm\!0.02$ for 30:1 PPO and $\beta\!=\!0.24\!\pm\!0.02$ for 6:1 PPO. For
the 15:1 composition, determination of the stretching parameter is somewhat ambiguous due
to the extreme nonexponentiality. However, for all studied temperatures, a reasonable
interpolation of the data can be obtained with $\beta\!=\!0.2$, as will be discussed in
more detail in Sec.\ \ref{15}. Hence, the values of the stretching parameters confirm
that the nonexponentiality of the segmental motion depends on the salt concentration.

The mean correlation times can be obtained according to
$\langle\tau\rangle\!=\!(\tau/\beta)\,\Gamma (1/\beta)$ from the fit parameters, where
$\Gamma(x)$ is the $\Gamma$ function. In the following, we denote the mean time constants
resulting for $t_p\!=\!60\mathrm{\mu s}$ as mean jump correlation times
$\langle\tau_J\rangle$, since correlation functions for this relatively large evolution
time approximate $F_2^J(t_m)$, see Sec.~\ref{TP}. The mean jump correlation times for
PPO, 30:1 PPO and 6:1 PPO are shown in Fig.\ \ref{TAU}. We see that, for
$T_g\!<\!T\!\lesssim\!1.1\,T_g$, the temperature dependence of $\langle\tau_J\rangle$ is
well described by an Arrhenius law. While we find similar activation energies
$E_a\!=\!2.7\!\pm\!0.2\mathrm{\,eV}$ and $E_a\!=\!2.4\!\pm\!0.2\mathrm{\,eV}$ for PPO and
30:1 PPO, respectively, there is a higher value $E_a\!=\!4.1\!\pm\!0.2\mathrm{\,eV}$ for
6:1 PPO. The higher activation energy for the latter composition is consistent with the
slowing down of the segmental motion for high salt concentrations. In addition, we see in
Fig.\ \ref{TAU} that the correlation times are independent of the experimental setup, as
one would expect.

To obtain an estimate of the "NMR glass transition temperature", we extrapolate the
Arrhenius temperature dependence of the mean jump correlation time and determine the
temperatures at which $\langle\tau_J\rangle\!=\!100\mathrm{\,s}$.  We find temperatures
$205\mathrm{\,K}$ for PPO ($T_g\!=\!203\mathrm{\,K}$), $208\mathrm{\,K}$ for 30:1 PPO
($T_g\!=\!203\mathrm{\,K}$) and $283\mathrm{\,K}$ for 6:1 PPO
($T_g\!=\!272\mathrm{\,K}$), which are in reasonable agreement with the calorimetrically
determined values of $T_g$. Thus, the slowing down of the segmental motion upon addition
of salt reflects the shift of the glass transition temperature.

Next, we focus on $F_2^{SS}(t_m;t_p\!=\!2\mathrm{\,\mu s})\! \approx\!F_2(t_m)$. Since
correction for SLR is not possible for sin-sin correlation functions, see
Sec.~\ref{THEO}, this analysis is restricted to relatively high temperatures at which
segmental motion is faster than SLR. In Fig.\ \ref{TAU}, we include the mean correlation
times $\langle\tau_C\rangle$, as obtained from fits of
$F_2^{SS}(t_m;t_p\!=\!2\mathrm{\,\mu s})$ to Eq.~(\ref{KWW}). We see that
$\langle\tau_C\rangle$ and $\langle\tau_J\rangle$ exhibit a comparable temperature
dependence. For all compositions, $\langle\tau_C\rangle$ is about one order of magnitude
longer than $\langle\tau_J\rangle$. Thus, a similar ratio
$\langle\tau_J\rangle/\langle\tau_C\rangle\!\approx\!0.1$ implies that the presence of
salt does not affect the jump angles of the segmental motion. Using this ratio in Eq.\
(\ref{gamma}), we obtain a mean jump angle $\gamma\!=\!15^{\circ}$. Hence, the segmental
motion involves rotational jumps about small angles both in the absence and in the
presence of salt. In Sec.\ \ref{TP}, the dependence of the mean correlation time on the
evolution time, $\langle\tau (t_p)\rangle$, will be studied in more detail.

In dielectric spectroscopy, it was found that a Vogel-Fulcher behavior describes the
temperature dependence of the $\alpha$ process of PPO in the presence and in the absence
of salt.\cite{Furukawa,Leon,Johari} Thus, on first glance, it is surprising that the
correlation times follow an Arrhenius law in our case. For PPO, we show
$\langle\tau_J(T)\rangle$ and $\langle\tau_C(T)\rangle$ together with dielectric data
$\tau_{\alpha}(T)$\cite{Furukawa,Leon} in Fig.\ \ref{TAU}(b). The time constants are
plotted on a reduced $T_g/T$ scale to remove effects resulting from the somewhat
different glass transition temperatures of the studied polymers. Considering the error
resulting from this scaling, we conclude that our results are in reasonable agreement
with the literature data. In particular, all the curves exhibit, if at all, a very small
curvature in the temperature range $0.9\!<\!T_g/T\!<\!1$, rationalizing that an Arrhenius
law enables a good description of the temperature dependence in this regime. At higher
temperatures, the dielectric data show a pronounced curvature in the Arrhenius
representation, as described by the Vogel-Fulcher behavior. In our case, the observation
of unphysical pre-exponential factors $\tau_0\!=\!10^{-56}\!-\!10^{-71}\mathrm{\,s}$
implies a breakdown of the Arrhenius law at higher temperatures. Consistently, a previous
SE study on the $\alpha$ process of a polymer melt reported prefactors
$\tau_0\!\approx\!10^{-77}\mathrm{\,s}$ from Arrhenius interpolations of temperature
dependent correlation times.\cite{Tracht} Hence, SE spectroscopy yields \emph{apparent}
activation energies $E_a$, characterizing the strong temperature dependence of the
segmental motion close to $T_g$, while deviations from an Arrhenius law become evident
when following $\tau_{\alpha}(T)$ to higher temperatures.

\subsection{$^2$H NMR correlation functions for various relaxation delays and evolution times}\label{TP}

\begin{figure}
\includegraphics[angle=270,width=8.2cm]{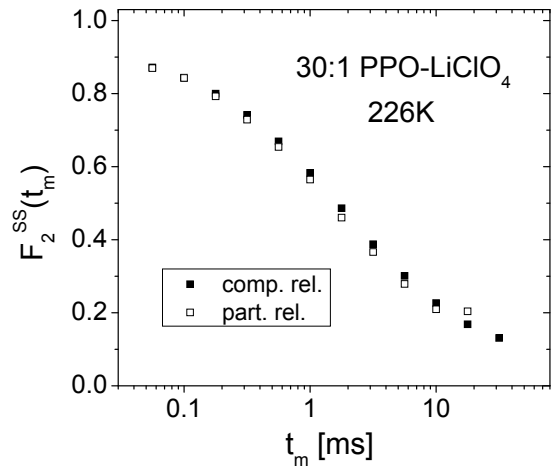}
\caption{Correlation functions $F_2^{SS}(t_m;t_p\!=\!2\mathrm{\,\mu s})$ for 30:1
PPO-LiClO$_4$ at $T\!=\!226\mathrm{\,K}$. The measurements were performed for complete
recovery ($t_R\!=\!1\mathrm{\,s}\!\gg\!T_{1,B}$) and partial recovery
($t_R\!=\!3\mathrm{\,ms}\!\ll\!T_{1,B}$) of the magnetization after
saturation.}\label{CH23}
\end{figure}

Next, we study whether different structural units of the PPO molecule show
distinguishable reorientation behavior. Exploiting that $T_{1,M}\!\ll\!T_{1,B}$ for PPO
and 30:1 PPO at $T\!\approx\!T_g$,\cite{Vogel} we record correlation functions for
different relaxation delays $t_R$ between the saturation of the magnetization and the
start of the SE pulse sequence. Then, the contribution of the M deuterons can be singled
out in partially relaxed experiments, i.e., for $t_R\!\approx\!T_{1,M}\!\ll\!T_{1,B}$,
while, of course, both deuteron species will contribute to the correlation functions if
we wait until the recovery of the magnetization is complete.

In Fig.~\ref{CH23}, we display completely relaxed and partially relaxed correlation
functions for 30:1 PPO at $T\!=\!226\mathrm{\,K}$. Evidently,
$F_2^{SS}(t_m;t_p\!=\!2\mathrm{\,\mu s})$ is independent of the relaxation delay $t_R$,
indicating that the ''relevant'' rotational jumps of the B and the M deuterons occur on
essentially the same time scale. Here, the term ''relevant'' is used to denote the
situation that the threefold jumps of the methyl group are not probed in the present
approach. Performing an analogous analysis for neat PPO, we arrive at the same
conclusion. Partially and completely relaxed correlation functions for larger values of
$t_p$ will be compared below.

\begin{figure}
\includegraphics[angle=270,width=8.2cm]{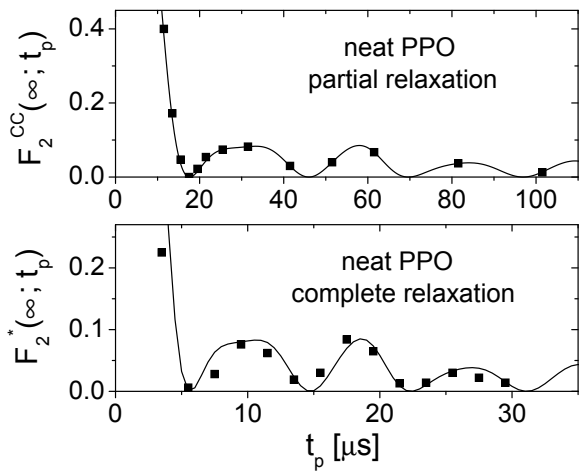}
\caption{Residual correlations from interpolations of cos-cos correlation functions for
PPO at $T\!=\!222\mathrm{\,K}$ with Eq.~(\ref{KWW}). The top and bottom panel show data
obtained for partial and complete recovery of the magnetization, respectively. The lines
are theoretical values $z_y^{CC}/a_y^{CC}$ for
$\delta_M\!=\!2\!\pi\times\!39\mathrm{\,kHz}$ (top panel) and
$\delta_B\!=\!2\!\pi\times\!119\mathrm{\,kHz}$ (bottom panel)}\label{ZEXP}
\end{figure}

In previous work on molecular and polymeric glass formers, it was demonstrated that more
detailed insights into the reorientational mechanisms are possible when
$F_2^{CC}(t_m;t_p)$ is recorded for many evolution
times.\cite{Roessler,Geil,Hinze,BoHi,Tracht} For example, information about the shape of
the distribution of jump angles is available from the dependence of the mean correlation
time on the evolution time, $\langle\tau(t_p)\rangle$. Hence, it is desirable to apply
this approach also to polymer electrolytes PPO-LiClO$_4$. However, in our case, the
presence of two deuteron species hampers straightforward analysis of $F_2^{CC}(t_m;t_p)$
for an arbitrary value of $t_p$, see Sec.~\ref{THEO}. In the following, we test three
approaches aiming to extract $\langle\tau(t_p)\rangle$ in such situation.

In approach (A), we measure partially relaxed correlation functions $F_2^{CC}(t_m;t_p)$.
Then, the signal solely results from the M deuterons and correction for SLR is possible
for all values of $t_p$ when the data is divided by the relaxation function
$R_{1,M}(t_m)$, which can be obtained from SLR measurements. Thus, it is possible to
determine $\langle\tau(t_p)\rangle$ from fits of the corrected data to Eq.\ (\ref{KWW}).
To demonstrate that this approach indeed allows us to single out the contribution of the
M deuterons, we show experimental and theoretical values of the residual correlation
$F_2^{CC}(\infty;t_p)$ as a function of $t_p$ in Fig.\ \ref{ZEXP}. The experimental
values result from interpolations of the partially relaxed correlation functions
$F_2^{CC}(t_m;t_p)$ for PPO at $T\!=\!222\mathrm{\,K}$. The theoretical values were
obtained by calculating $z_M^{CC}(t_p)/a_M^{CC}(t_p)$ for
$\delta_M\!=\!2\pi\times39\mathrm{\,kHz}$.\cite{Vogel} Obviously, the experimental and
the theoretical curves $F_2^{CC}(\infty;t_p)$ nicely agree, when we shift the measured
data by $1.5\mathrm{\,\mu s}$ to longer evolution times. This shift reflects the fact
that, due to finite pulse lengths, the effective evolution times are
$1\!-\!2\mathrm{\,\mu s}$ longer than the set values, as was reported in previous
work.\cite{BoHi} The good agreement of experimental and theoretical residual correlations
indicates that approach (A)  allows us to study the reorientational mechanisms of the
methyl group axes. However, this approach is restricted to PPO and 30:1 PPO since
$T_{1,M}$ and $T_{1,B}$ differ less for higher salt concentrations,\cite{Vogel} hampering
a suppression of contributions from the B deuterons in partially relaxed experiments.

In approaches (B) and (C), we focus on completely relaxed correlation functions. Then, we
can include 6:1 PPO, for which the salt should have the strongest effect on the segmental
motion. In both approaches, the measured decays for different evolution times are divided
by the corresponding effective SLR functions, which are determined in concomitant SLR
measurements, as is described in the Appendix. To emphasize that correction for SLR is
not successful when $z_B^{CC}(t_p),z_M^{CC}(t_p)\!\neq\!0$, we denote the resulting data
as $F_2^{*}(t_m;t_p)$, see Eq.~(\ref{FTP}).

In approach (B), we extract $\langle\tau(t_p)\rangle$ from interpolations of
$F_2^{*}(t_m;t_p)$ with Eq.~(\ref{KWW}). In doing so, we disregard that the residual
correlation depends on the mixing time at $t_m\!\approx\!T_{1,M}$, see Appendix. In the
limit $t_m\!\rightarrow\!\infty$, a residual correlation
$F_2^{*}(\infty;t_p)\!=\!z_B^{CC}(t_p)/a_B^{CC}(t_p)$ is expected based on
Eq.~(\ref{REST}). Figure~\ref{ZEXP} shows the residual correlation obtained from
completely relaxed correlation functions $F_2^{*}(t_m;t_p)$ for PPO at
$T\!=\!222\mathrm{\,K}$ together with the theoretical curve $z_B^{CC}(t_p)/a_B^{CC}(t_p)$
for $\delta_B\!=\!2\pi\times119\mathrm{\,kHz}$.\cite{Vogel} Shifting the measured points
again by $1.5\mathrm{\,\mu s}$, a good agreement of the experimental and the theoretical
data is obtained, confirming our expectations for the residual correlation. Therefore, we
use the theoretical values of $F_2^{*}(\infty;t_p)$ in a second round of fitting to
reduce the number of free parameters and the scattering of the data.

\begin{figure}
\includegraphics[angle=270,width=8.2cm]{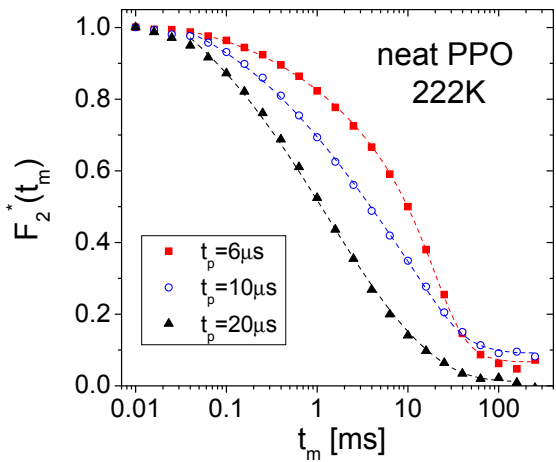}
\caption{$F_2^{*}(t_m;t_p)$ for PPO at $T\!=\!222\mathrm{\,K}$. The evolution times $t_p$
are indicated. The dashed lines are fits with Eq. (\ref{FTP}). See text for
details.}\label{FITS}
\end{figure}

In approach (C), we employ Eq.\ (\ref{FTP}) to fit $F_2^{*}(t_m;t_p)$. In doing so, we
use the theoretical values of $z_y^{CC}(t_p)$ and $a_y^{CC}(t_p)$ together with the
relaxation functions $R_{1,y}$ and $R_{2,y}$ from independent SLR measurements, see
Appendix. Considering our result that the relevant rotational jumps of the B and the M
deuterons occur on very similar time scales, see Fig.~\ref{CH23}, we utilize the same
correlation function $\phi(t_m;t_p)$ for both deuteron species. Assuming a KWW function,
we have two free fit parameters, $\tau$ and $\beta$. In Fig.\ \ref{FITS}, we show
examples of $F_2^{*}(t_m;t_p)$ for PPO at $T\!=\!222\mathrm{\,K}$. Evidently, approach
(C) enables a good description of the data for all values of $t_p$.

\begin{figure}
\includegraphics[angle=0,width=8.2cm]{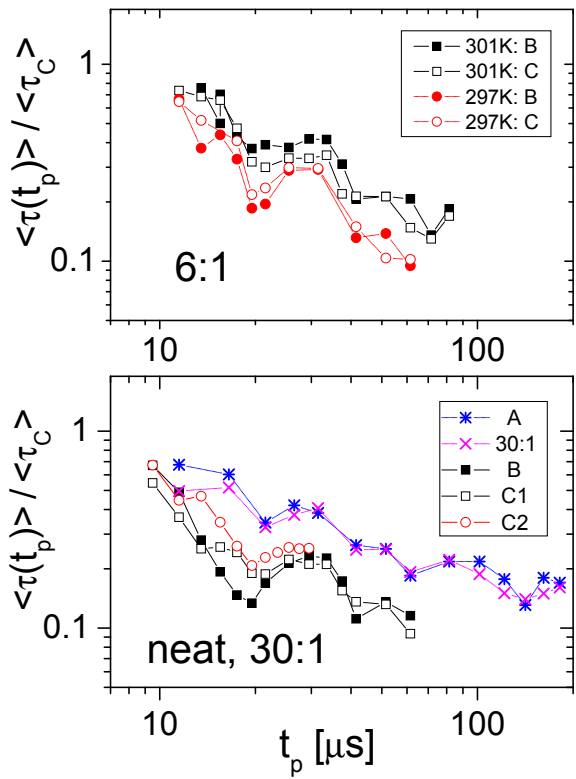}
\caption{Scaled mean correlation times $\langle\tau(t_p)\rangle/\langle\tau_C\rangle$ for
PPO at $T\!=\!222\mathrm{\,K}$ and 6:1 PPO-LiClO$_4$ at $T\!=\!297\mathrm{\,K}$ and
$T\!=\!301\mathrm{\,K}$. Results from approach (B) and approach (C) are included. For
PPO, we compare data from two independent measurements (C1,C2). In addition, we show the
scaled mean correlation times from partially relaxed experiments for PPO at
$T\!=\!222\mathrm{\,K}$ and for 30:1 PPO-LiClO$_4$ at $T\!=\!226\mathrm{\,K}$
($t_R\!=\!3\mathrm{\,ms}$), which were obtained using approach (A). See text for
details.}\label{TAUTP}
\end{figure}

In Fig.\ \ref{TAUTP}, we present mean correlation times $\langle\tau(t_p)\rangle$
resulting from the described approaches for PPO, 30:1 PPO and 6:1 PPO. Strictly speaking,
the scaled data
$\widetilde\tau(t_p)\!\equiv\!\langle\tau(t_p)\rangle/\langle\tau_C\rangle$ are shown for
the sake of comparison. Inspecting the results of the partially relaxed experiments, it
is evident that the rotational jumps in PPO and 30:1 PPO are characterized by very
similar curves $\widetilde\tau(t_p)$. Hence, the threefold methyl group axes show
essentially the same reorientational mechanism in both materials. In particular, for both
compositions, the dependence on the evolution time is weak for $t_p\!>\!60\mathrm{\,\mu
s}$, indicating that the jump correlation function $F_2^J(t_m)$ is approximately
measured. Thus, $\widetilde\tau(t_p\!>\!60\mathrm{\,\mu s})$ provides an estimate for
$\langle \tau_J\rangle/\langle\tau_C\rangle$. Using a typical value
$\widetilde\tau\!=\!0.15$, Eq.\ (\ref{gamma}) yields a mean jump angle
$\gamma\!=\!18^\circ$ for the rotational jumps of the methyl-group axes.

Next, we focus on the outcome of the completely relaxed experiments in Fig.~\ref{TAUTP}.
Although approaches (B) and (C) give somewhat different results, the overall shape of the
respective curves $\widetilde\tau(t_p)$ is still comparable. In particular, the findings
for sufficiently large evolution times are in good agreement due to small values of the
residual correlations $z_y^x(t_p)$, see Fig.~\ref{AZ}. We conclude that semiquantitative
information about the reorientational mechanism is available in our case of two deuteron
species. While the data allows us to compare the results for different salt
concentrations and to extract mean jump angles, we refrain from determination of the
distribution of jump angles from further analysis of $\widetilde\tau(t_p)$, since the
error is too big to resolve details of the shape of this distribution. We estimate the
data to be correct within a factor of two. The rough agreement of the curves
$\widetilde\tau(t_p)$ for PPO and 6:1 PPO nevertheless indicates that the presence of
salt, if at all, has a weak effect on the reorientational mechanism of the polymer
segments, in harmony with the results in Fig.\ \ref{TAU}. In Fig.~\ref{TAUTP}, we further
see that $\widetilde\tau(t_p)$ weakly depends on $t_p$ for $t_p\!>\!40\mathrm{\,\mu s}$,
justifying our argument $F_2^{CC}(t_m;t_p\!=\!60\mathrm{\,\mu s})\!\approx\!F_2^J(t_m)$.
For both materials, we find $\widetilde\tau(t_p\!>\!40\mathrm{\,\mu s})\!=\!0.1\!-\!0.2$,
corresponding to mean jump angles $\gamma\!=\!15\!-\!21^{\circ}$.

\begin{figure}
\includegraphics[angle=270,width=7.5cm]{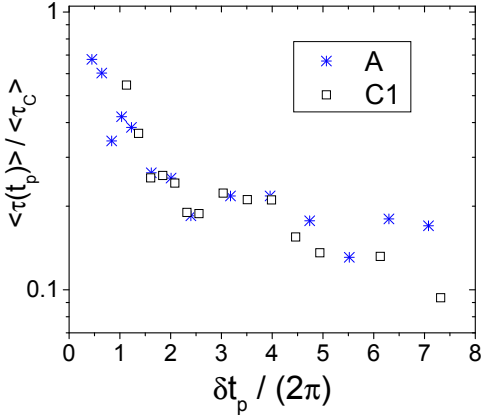}
\caption{Scaled mean correlation times $\langle\tau(t_p)\rangle/\langle\tau_C\rangle$
from partially relaxed (A) and completely relaxed (C1) correlation functions for PPO at
$T\!=\!222\mathrm{\,K}$. The data is the same as in Fig.\ \ref{TAUTP}, but it is now
shown on a reduced $\delta t_p$ scale. We used
$\delta_M\!=\!2\!\pi\times\!39\mathrm{\,kHz}$ and
$\delta_B\!=\!2\!\pi\times\!119\mathrm{\,kHz}$ for the partially and the completely
relaxed data, respectively.}\label{TPSCAL}
\end{figure}

In Fig.~\ref{TPSCAL}, we replot the results from the partially and the completely relaxed
experiments for PPO at $T\!=\!222\mathrm{\,K}$ on a reduced $\delta t_p$ scale. In this
way, we can alleviate effects due to the difference of the anisotropy parameters
$\delta_M\!\neq\!\delta_B$. The good agreement of the completely and the partially
relaxed data indicates that the relevant rotational jumps of the B and the M deuterons
are similar not only with respect to the time scale, see Fig.\ \ref{CH23}, but also with
respect to the reorientational mechanism. Some deviations are expected since M and B
deuterons contribute to the completely relaxed data and, hence, complete elimination of
effects due to the different anisotropy parameters is not possible.

\subsection{$^2$H NMR and $^7$Li NMR correlation functions for 15:1 PPO-LiClO$_4$}\label{15}

\begin{figure}
\includegraphics[angle=270,width=8.7cm]{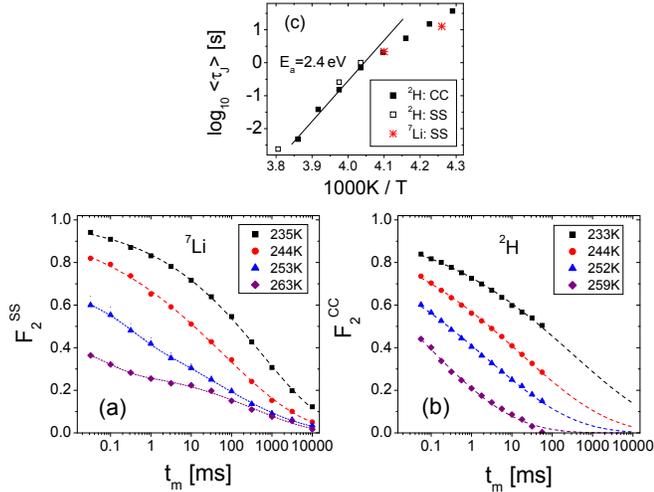}
\caption{Correlation functions for 15:1 PPO-LiClO$_4$ at the indicated temperatures: (a)
$F_2^{SS}(t_m;t_p\!=\!60\mathrm{\,\mu s})$ from $^7$Li NMR and (b)
$F_2^{CC}(t_m;t_p\!=\!60\mathrm{\,\mu s})$ from $^2$H NMR. The dashed lines are fits with
a KWW function, see Eq.\ (\ref{KWW}), where $C\!=\!0.00$ and $C\!=\!0.01$ for $^7$Li and
$^2$H NMR, respectively. The dotted lines are interpolations of the $^7$Li NMR data for
$T\!=\!253\mathrm{\,K}$ and $T\!=\!263\mathrm{\,K}$ with a two-step decay. (c) Mean
correlation times $\langle\tau_J\rangle$ determined based on the KWW fits of the $^2$H
and $^7$Li NMR results. We also include time constants obtained from measurements of
$F_2^{SS}(t_m;t_p\!=\!60\mathrm{\,\mu s})$ in $^2$H NMR.}\label{LI}
\end{figure}

In Fig.~\ref{LI}, we compare $^2$H and $^7$Li NMR correlation functions for 15:1 PPO. It
is evident that the correlation functions for both nuclei decay on similar time scales,
suggesting a coupling of ion and polymer dynamics. In $^2$H NMR, reasonable
interpolations of $F_2^{CC}(t_m;t_p\!=\!60\mathrm{\,\mu s})$ are obtained for all studied
temperatures when we use a KWW function with $\beta\!=\!0.2$. However, closer inspection
of the data reveals some deviations from a KWW behavior. Considering that a coexistence
of fast and slow polymer dynamics was reported for compositions of about 15:1
PPO-LiClO$_4$,\cite{Furukawa,Bergman} one may speculate that a fit with two Kohlrausch
functions gives a better description of the $^2$H NMR correlation functions. However, the
limited time window does not allow us an unambiguous decision. In any case, the shape of
the $^2$H NMR correlation function shows that if fast and slow polymer dynamics exist the
corresponding rate distributions will overlap in large parts so that there is no true
bimodality. The $^7$Li NMR correlation functions $F_2^{SS}(t_m;t_p\!=\!60\mathrm{\,\mu
s})$ for $T\!<\!250\mathrm{\,K}$ are well described by a KWW function with
$\beta\!\approx\!0.24$. At higher temperatures, two-step decays are evident, suggesting
the presence of two distinguishable lithium species. However, since the interpretation of
$^7$Li NMR data for polymer electrolytes has not yet been established, this tentative
speculation deserves a careful check in future work, see Sec.~\ref{NMR}.

The mean time constants resulting from the KWW fits of the $^2$H and $^7$Li NMR
correlation functions are compiled in Fig.\ \ref{LI}(c). At variance with the results for
the other compositions, the mean jump correlation times $\langle \tau_J \rangle$
extracted from the $^2$H NMR data for 15:1 PPO do not follow an Arrhenius law. For
$T\!>\!245\mathrm{\,K}$, the temperature dependence is described by
$E_a\!=\!2.4\mathrm{\,eV}$, which is consistent with the apparent activation energies for
PPO and 30:1 PPO, see Fig.\ \ref{TAU}. However, the variation of $\langle \tau_J \rangle$
with temperature is weaker at lower temperatures. There are two possible explanations for
the latter behavior. On the one hand, it is possible that a description using a single
KWW function is not appropriate due to an existence of fast and slow polymer dynamics. On
the other hand, it was demonstrated that the decay times of correlation functions do no
longer reflect the shift of a distribution of correlation times when this distribution is
much broader than the experimental time window,\cite{Vogel_1,Vogel_2} as one may expect
for 15:1 PPO.\cite{Vogel} Rather, due to interference with the experimental time window,
the decay times feign a weaker temperature dependence,\cite{Vogel_3} as is observed in
our case. The mean correlation times resulting from the $^7$Li NMR experiments are
comparable to the corresponding $^2$H NMR data. This finding suggests a coupling of ion
and polymer dynamics, as was inferred from visual inspection of the correlation
functions. However, we again emphasize that, in studies on the ion dynamics in polymer
electrolytes, the interpretation of $^7$Li NMR experiments is not yet clear.

\section{Summary and Conclusion}\label{Discussion}

We demonstrated that $^2$H NMR SE spectroscopy provides straightforward access to both
the time scale and the geometry of slow polymer segmental motion in polymer electrolytes.
Here, we exploited these capabilities for an investigation of PPO, 30:1 PPO-LiClO$_4$,
15:1 PPO-LiClO$_4$ and 6:1 PPO-LiClO$_4$. Comparison of the results for the different
salt concentrations gives clear evidence that addition of salt leads to a strong slowing
down of the polymer dynamics, while the geometry of the segmental motion is hardly
affected.

The dependence of the $^2$H NMR correlation functions on the evolution time yields
quantitative information about the geometry of the segmental
reorientation.\cite{Fujara_1,Roessler} This possibility results from the fact that the
evolution time has the meaning of a geometrical filter. The jump angles of rotational
jumps can be determined from the variation of the mean correlation time with the
evolution time, $\langle\tau(t_p)\rangle$.\cite{Geil,Hinze,BoHi,Tracht} For PPO and 6:1
PPO, we found comparable curves $\langle\tau(t_p)\rangle$, indicating that the polymer
segments exhibit similar jump angle distributions in the presence and in the absence of
salt. Furthermore, we exploited that the mean jump angle can be determined from the ratio
of the jump correlation time $\langle\tau_J\rangle$ and the overall correlation time
$\langle\tau_C\rangle$, which can be extracted from the $^2$H NMR correlation functions
in the limits $t_p\!\rightarrow\!\infty$ and $t_p\!\rightarrow\!0$, respectively. For
PPO, 30:1 PPO and 6:1 PPO, a ratio of about 0.1 corresponds to a mean jump angle of ca.\
15$^{\circ}$. In the case of PPO, this value is consistent with results for various neat
polymers,\cite{Spiess,Wefing,Kaufmann,Pschorn,Schaefer,Tracht} suggesting that small
angle rotational jumps are typical of polymer segmental motion close to $T_g$. For the
polymer electrolytes, the agreement of the mean jump angles confirms that the
reorientational mechanism is essentially independent of the salt concentration.

In general, \emph{trans-gauche} conformational transitions are attributed to the $\alpha$
process, implying the existence of large angle jumps. Thus, it may be surprising that the
present and previous NMR studies on polymer melts and polymer electrolytes find small
jump angles of $10\!-\!40^{\circ}$. To rationalize the latter behavior, the cooperativity
of the $\alpha$ process was taken into account in the literature.\cite{Tracht}
Specifically, it was argued that concerted conformational transitions of neighboring
polymer segments lead to jump angles that are not well-defined and smaller than the
angles typical of conformational transitions.

Comparing the correlation times for the different mixtures PPO-LiClO$_4$, we quantified
the slowing down of the segmental motion as a consequence of the presence of salt. For
PPO, the mean jump correlation time amounts to $\langle\tau_J\rangle\!=\!1\mathrm{\,s}$
at $T\!=\!212\mathrm{\,K}$, whereas, for the highest salt concentration, 6:1 PPO, it has
this value at $T\!=\!292\mathrm{\,K}$. Considering the strong temperature dependence of
polymer dynamics close to $T_g$, the difference $\Delta T\!=\!\mathrm{80\,K}$ shows that
addition of salt results in an enormous slowing down of the polymer segmental motion.
This finding is consistent with both pronounced effects on the glass transition
temperature and results from previous work.\cite{Carlsson_2,Johari}

In the studied temperature ranges of $T_g\!<\!T\!\lesssim\!1.1\,T_g$, the temperature
dependence of the mean jump correlation time $\langle\tau_J\rangle$ follows an Arrhenius
law. While we observed apparent activation energies $E_a\!\approx\!2.5\mathrm{\,eV}$ for
the neat polymer and the moderate salt concentrations, a higher value
$E_a\!\approx\!4.1\mathrm{\,eV}$ characterizes the segmental motion in 6:1 PPO. Likewise,
the mean correlation time $\langle\tau_C\rangle$, characterizing the rotational
correlation function of the second Legendre polynomial, shows a substantially stronger
temperature dependence for 6:1 PPO than for PPO. Hence, the slowing down of the polymer
dynamics for high salt concentrations is also reflected in higher apparent activation
energies close to $T_g$. For all compositions, the pre-exponential factors of the
Arrhenius laws have unphysical values, $\tau_0\!=\!10^{-56}\!-\!10^{-71}\mathrm{\,s}$,
indicating that there is no simple thermally activated process over a single barrier.
Consistently, a Vogel-Fulcher behavior can be used to describe the temperature dependence
of the $\alpha$ process in a broader temperature
range.\cite{Furukawa,Johari,Bergman,Leon}

The $^2$H NMR correlation functions for mixtures PPO-LiClO$_4$ decay in a strongly
nonexponential manner, where the stretching depends on the salt concentration. For the
correlation function $F_2^{CC}(t_m;t_p\!=\!60\mathrm{\,\mu s})$, we obtained stretching
parameters $\beta\!\approx\!0.33$ for PPO, $\beta\!\approx\!0.3$ for 30:1 PPO,
$\beta\!\approx\!0.2$ for 15:1 PPO and $\beta\!\approx\!0.24$ for 6:1 PPO, which show no
systematic variation with temperature in the range $T_g\!<\!T\!\lesssim\!1.1\,T_g$. Thus,
the nonexponentiality of the segmental motion is most prominent for intermediate salt
concentrations of about 15:1, while it is reduced for both low and high salt content.
Likewise, previous work reported that polymer dynamics is more nonexponential in the
presence than in the absence of salt.\cite{Carlsson_2,Torell} We conclude that the
additional structural heterogeneities imposed by the addition of salt have a strong
effect on the structural relaxation of the polymer melt.

Comparison of $^2$H NMR correlation functions for partial and complete recovery of the
longitudinal magnetization provided insights into the reorientation of different
structural units. With the exception of methyl group rotation, we found no evidence for
diverse dynamical behaviors of different structural units. Strictly speaking, the C-D
bonds of the ethylene groups and the threefold methyl group axes show comparable
rotational jumps with respect to both time scale and jump angles.

All these results of the present $^2$H NMR SE study are consistent with findings of our
previous $^2$H NMR line-shape analysis for polymer electrolytes
PPO-LiClO$_4$.\cite{Vogel} Specifically, the $^2$H NMR spectra gave evidence that the
presence of salt slows down the segmental motion. In addition, the $^2$H NMR spectra
indicated that, in large parts, the pronounced nonexponentiality of the polymer dynamics
results from the existence of a broad distribution of correlation times $G(\log \tau)$.
Then, the stretching parameters $\beta$ are a measure of the decadic width $\Delta$ of
this distribution. In detail, the width is given by
$\Delta\!\approx\!1.14(0.93/\beta\!+\!0.06)$.\cite{BoBe} Based on this expression, we
obtain widths of $\Delta\!=\!3.3$ and $\Delta\!=\!5.4$ orders of magnitude from the
stretching parameters for PPO and 15:1 PPO, respectively. The latter result is in nice
agreement with findings from our $^2$H NMR line-shape analysis for the 15:1
composition.\cite{Vogel} There, the spectra were described using a logarithmic Gaussian
distribution $G(\log \tau)$ with a full width at half maximum of 5-6 orders of magnitude.
We conclude that both $^2$H NMR techniques yield a coherent picture of the segmental
motion in polymer electrolytes. However, they highlight different aspects of the dynamics
and, hence, a combined approach is particularly useful. While $^2$H NMR SE spectroscopy
is a powerful tool to quantify the time scale and the geometry of the segmental motion,
$^2$H NMR spectra allow one to demonstrate the existence of dynamical heterogeneities
provided the distribution $G(\log \tau)$ is sufficiently broad.

Having established that presence of salt slows down the segmental motion, one expects
that the shape of the distribution of correlation times governing the polymer dynamics
reflects features of the salt distribution in polymer electrolytes PPO-LiClO$_4$. We
observed that the distribution $G(\log \tau)$ is broadest for the 15:1 composition,
implying a strong variation in the local salt concentration for this intermediate
composition, whereas the spatial distribution of the salt is more homogeneous for both
high and low salt content. In the literature, it was proposed that there is liquid-liquid
phase separation into salt-rich and salt-depleted regions for intermediate salt
concentrations.\cite{Vachon_1,Vachon_2,Furukawa,Bergman} Provided the rate of the
segmental motion is a measure of the local salt concentration, the existence of
well-defined microphases should lead to a bimodal distribution of correlation times for
the polymer dynamics and, consequently, to a two-step decay of the $^2$H NMR correlation
functions. Although there are some indications for the presence of a two-step decay, a
KWW function still provides a satisfactory interpolation of the experimental data for
15:1 PPO-LiClO$_4$. Hence, our results are at variance with a bimodal, discontinuous
distribution of correlation times and, thus, of local salt concentrations unless the
contributions from salt-rich and salt-depleted regions overlap in large parts, while they
are consistent with the existence of a continuous $G(\log \tau)$. In particular, the
$^2$H NMR spectra and correlation functions for 15:1 PPO are in harmony with the picture
that short ranged fluctuations of the local salt concentration lead to a large structural
diversity and, thus, to a broad continuous rate distribution for the segmental motion.
For example, our results are consistent with a reverse Monte-Carlo model of 16:1
PPO-LiClO$_4$, which features salt-rich and salt-depleted regions on a short length scale
of about $\mathrm{10\,\AA}$,\cite{Carlsson_3} corresponding to about two monomeric units.

In addition, we performed $^7$Li NMR SE experiments to investigate the lithium ionic
motion in 15:1 PPO-LiClO$_4$. For polymer electrolytes, a strong coupling of ion and
polymer dynamics was proposed in the literature.\cite{Armand,Ratner} Consistently, we
found nonexponential $^7$Li NMR correlation functions, which agree roughly with the
corresponding $^2$H NMR data. Hence, our results are not at variance with the common
picture of charge transport in polymer electrolytes. However, we emphasize that the
present findings cannot confirm a coupling of ion and polymer dynamics beyond doubt,
since the interpretation of $^7$Li NMR studies on lithium ionic motion in polymer
electrolytes needs to be further clarified. Specifically, it is not clear to which extent
the time dependence of the $^7$Li NMR resonance frequency reflects the dynamics of the
ion and the rearrangement of neighboring polymer segments, respectively.

From the viewpoint of NMR methodology, we investigated whether $^2$H NMR SE spectroscopy
provides well defined information when there are contributions from two or more deuteron
species, exhibiting different relaxation behaviors. We showed that problems can arise
when at least one of the deuteron species exhibits fast SLR so that the decay of the
$^2$H NMR correlation functions due to relaxation interferes with that due to molecular
dynamics. It was demonstrated that a successful correction for SLR requires that the
different deuteron species show similar molecular dynamics. Moreover, it relies on the
choice of appropriate evolution times $t_p$. While the effects of SLR can be eliminated
for evolution times $t_p$, for which the residual correlations $z(t_p)$ vanish for all
deuteron species, correction for relaxation effects is not possible in the general case.
Then, a more elaborate analysis can yield some insights, however the experimental error
increases.

\section{Appendix}

Here, we discuss whether relaxation effects can be eliminated when the experimental data
are divided by an effective SLR function, see Eq.\ (\ref{FTP}). We assume that the same
correlation function $\phi(t_m;t_p)$ characterizes the relevant rotational jumps of the B
and the M deuterons, as is suggested by experimental findings, see Fig.\ \ref{CH23}.
Then, short calculation shows that $F_2^{*}(t_m;t_p)\!=\!\phi(t_m;t_p)$ for
$z_M^{CC}(t_p)\!=\!z_B^{CC}(t_p)\!=\!0$. Consequently, the cos-cos correlation functions
provide straightforward access to the molecular reorientation when we use evolution
times, for which the residual correlations of both deuteron species vanish. However, such
correction for SLR is not possible for an arbitrary value of $t_p$. To illustrate the
effects it suffices to consider the case $R_{2,B}(2t_p)\!=\!R_{2,M}(2t_p)$. First, it is
instructive to study the residual correlation of $F_2^{*}(t_m;t_p)$. For
$t_m\!\gg\!\tau$, we obtain from Eq.\ (\ref{R})
\begin{equation}\label{REST}
F_2^{*}(t_{m}\!\gg\!\tau;t_p)=\frac{z_M^{CC}(t_p) \rho_1
(t_m)+z_B^{CC}(t_p)}{a_M^{CC}(t_p) \rho_1(t_m)+a_B^{CC}(t_p)}
\end{equation}
where $\rho_1(t_m)\!=\!R_{1,M}(t_m)/R_{1,B}(t_m)$. In our case $T_{1,M}\!\ll\!T_{1,B}$,
this function decays from $\rho_1(t_m\!\ll\!T_{1,M})\!=\!1$ to
$\rho_1(t_m\!\gg\!T_{1,M})\!=\!0$ and, thus, the residual correlation shows a crossover
from $[z_M\!+\!z_B]/[a_M\!+\!a_B]$ to $z_B/a_B$ at $t_m\!\approx\!T_{1,M}$, where we
skipped the superscripts. When the decays due to molecular dynamics and SLR are well
separated, i.e., for $\tau\!\ll\!T_{1,M}$,  this crossover is not relevant since the
analysis of the correlation functions can be restricted to mixing times
$t_m\!\ll\!T_{1,M}$. However, in the case $\tau\!\approx\!T_{1,M}$, which is often found
in the experiment due to broad distributions of correlation times, the crossover of the
residual correlation at $t_m\!\approx\!T_{1,M}$ interferes with the decay due to
molecular dynamics. Then, our approach does not enable an elimination of relaxation
effects.

\begin{figure}
\includegraphics[angle=0,width=7.5cm]{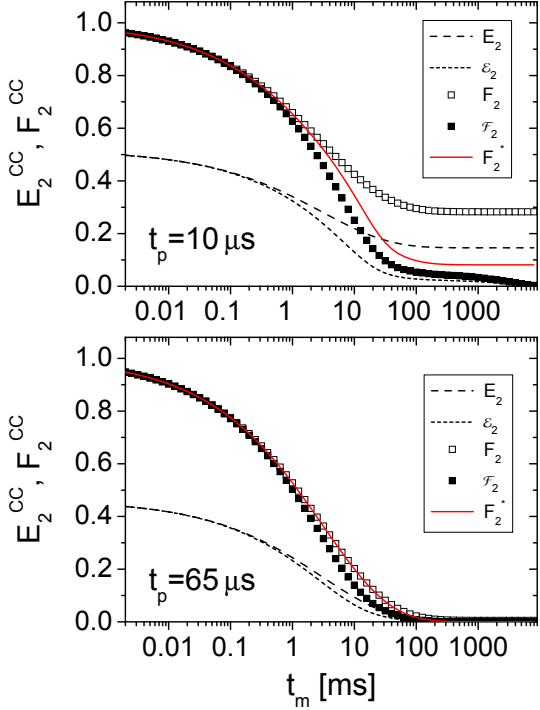}
\caption{Calculated cos-cos correlation functions for $t_p\!=\!10\mathrm{\,\mu s}$ and
$t_p\!=\!65\mathrm{\,\mu s}$. For the calculations, we utilized Eqs.\ (\ref{E}) and
(\ref{R}), where $\phi_y(t_m;t_p)\!=\!\exp[-(t_m/3\mathrm{\,ms})^{0.45}]$, $
R_{2,y}(2t_p)\!=\!1$, $R_{1,M}(t_m)\!=\!\exp(-t_m/10\mathrm{\,ms})$ and
$R_{1,B}(t_m)\!=\!\exp(-t_m/3\mathrm{\,s})$. The values of the initial and the final
correlations were computed for $\delta_B\!=\!2\pi\!\times\!120\mathrm{\,kHz}$ and
$\delta_M\!=\!2\pi\!\times\!40\mathrm{\,kHz}$ according to Eqs.\ (\ref{A}) and (\ref{Z}).
We show the correlation functions $E_2^{CC}(t_m;t_p)$ and ${\cal E}_2^{CC}(t_m;t_p)$,
differing due to the action of SLR, together with the normalized counterparts
$F_2^{CC}(t_m;t_p)$ and ${\cal F}_2^{CC}(t_m;t_p)$. Moreover, we include the correlation
functions $F_2^{*}(t_m;t_p)$ calculated according to Eq.\ (\ref{FTP}).}\label{MOD}
\end{figure}

To make the effects more clear, we show several calculated correlation functions in Fig.\
\ref{MOD}. For the calculation, we assume that the molecular rotational jumps are
described by a KWW function $\phi(t_m)\!=\!\exp[-(t_m/3\mathrm{ms})^{0.45}]$. Further,
$\delta_B\!=\!2\pi\!\times\!120\mathrm{\,kHz}$ and
$\delta_M\!=\!2\pi\!\times\!40\mathrm{\,kHz}$ are used to compute the initial and the
residual correlations. Utilizing these data, $E_2^{CC}$ is obtained from Eq.\ (\ref{E}).
To mimic the damping due to SLR, we calculate ${\cal E}_2^{CC}$ according to Eq.\
(\ref{R}). In doing so, we suppose exponential relaxation functions $R_{1,y}(t_m)$
characterized by $T_{1,M}\!=\!10\mathrm{\,ms}$ and $T_{1,B}\!=\!3\mathrm{\,s}$. Thus, the
decays due to dynamics and relaxation interfere for the M deuterons. In Fig.\ \ref{MOD},
the effect of SLR is evident from comparison of $E_2^{CC}$ and ${\cal E}_2^{CC}$.
Normalizing these data, we obtain $F_2^{CC}$ and ${\cal F}_2^{CC}$. Finally, Eq.\
(\ref{FTP}) is used to compute $F_2^{*}$. If our approach enables elimination of SLR
effects, $F_2^{*}(t_m)$ and $F_2^{CC}(t_m)$ will be identical.

In Fig.\ \ref{MOD}, we compare the various correlation functions for the evolution times
$t_p\!=\!10\mathrm{\,\mu s}$ and $t_p\!=\!65\mathrm{\,\mu s}$. For the latter evolution
time, the condition $z_B^{CC}\!=\!z_M^{CC}\!\approx\!0$ is met, see Fig.\ \ref{AZ}, so
that elimination of relaxation effects should be possible. This expectation is confirmed
by the nice agreement of $F_2^{*}(t_m)$ and $F_2^{CC}(t_m)$. By contrast, for
$t_p\!=\!10\mathrm{\,\mu s}$, we find $z_M^{CC}\!>\!z_B^{CC}\!>\!0$. Then, $F_2^{*}(t_m)$
clearly differs from $F_2^{CC}(t_m)$, indicating that correction for SLR is not
successful. In particular, $F_2^{*}(t_m)$ exhibits a residual correlation
$z_B^{CC}/a_B^{CC}$, which deviates substantially from the value resulting in the absence
of relaxation effects, see\ Eq.\ (\ref{REST}).

For an experimental determination of the effective SLR function, we measure the SLR using
the saturation recovery technique, where the longitudinal magnetization is read out with
a solid-echo so that SSR reduces the height of the solid-echo. For an evolution time
$t_p$ in the SE experiment, we perform a concomitant SLR experiment choosing the
solid-echo delay $t_e$ such that the total time intervals during which transversal
magnetization exists are equal in both measurement. In detail, we use $t_e\!=\!t_p$ or
$t_e\!=\!t_p\!+\!\Delta$, depending on whether a three-pulse or a four-pulse sequence is
applied in the SE experiment, see Sec.\ \ref{exp}. Then, in both the SLR and the SE
experiment, the contribution of each deuteron species is proportional to the
''effective'' magnetization $m_y R_{2,y}(2t_e)$. In particular, the recovery of the
longitudinal magnetization is the sum of the contributions $m_y R_{2,y}(2t_e)
R_{1,y}(t_m)$. Multiplication of these contributions with the respective theoretical
values of $a_y^{CC}(t_p)$ yields the effective SLR function.

At the studied temperatures, the recovery of the longitudinal magnetization occurs in two
steps, which can be attributed to the B and the M deuterons, respectively.\cite{Vogel}
Therefore, it is straightforward to determine the relaxation functions $R_{1,y}(t_m)$ and
the relative contributions of the two deuteron species. When we study the dependence of
$F_2^{CC}(t_m;t_p)$ on the evolution time and, hence, when we use various solid-echo
delays $t_e\!=\!t_p\!+\!\Delta$ in the concomitant SLR measurements, we find that
$R_{1,y}(t_m)$ does not exhibit a systematic variation with $t_e$, as one would expect.
Therefore, we use the average values for the calculation of the effective SLR functions.
However, the relative amplitudes of the two steps depend on the echo delay, indicating
that the B and the M deuterons show different SSR behavior. For PPO at
$T\!=\!222\mathrm{\,K}$, the contribution of the M deuterons increases from 55\% for
$t_e\!=\!25\mathrm{\mu s}$ to 65\% for $t_e\!=\!85\mathrm{\mu s}$. Hence, the M deuterons
exhibit slower SSR, consistent with a smaller anisotropy parameter $\delta$. When
extending the evolution time in SE experiments, this discrepancy in SSR means that the
relative contribution of the B deuterons decreases with respect to the theoretical value,
$a_B^x/(a_B^x\!+\!a_M^x)$.

\begin{acknowledgments}

We are grateful to J.\ Jacobsson for kindly providing us with deuterated PPO. Moreover,
we thank H.\ Eckert for allowing us to use his NMR laboratory, S.\ Faske for helping us
to prepare the samples and A.\ Heuer for valuable discussions. Finally, funding of the
Deutsche Forschungsgemeinschaft (DFG) through the Sonderforschungsbereich 458 is
gratefully acknowledged.
\end{acknowledgments}


\begin{thebibliography}{99}
\bibitem{34} Since we focus on nomalized correlation functions in the experimental part, we disregard the
usual factor $3/4$ in Eq.\ (\ref{SS}). This allows us a combined discussion of $E_2^{SS}$
and $E_2^{CC}$.
\end{thebibliography}
\end{document}